\documentclass[aps,11pt,letterpaper]{revtex4}
\usepackage{amsmath,amssymb,mathptmx,slashed,bm}
\usepackage[pdftex]{graphicx}
\newcommand{\Lx}{\left(}
\newcommand{\Rx}{\right)}
\newcommand{\LB}{\left[}
\newcommand{\RB}{\right]}

\newcommand{\ep}{{\varepsilon}}

\newcommand{\qcd} {{QCD}}
\newcommand{\qed} {{QED}}

\newcommand{\aspi} {{\Lx\frac{\alpha_s}{\pi}\Rx}}
\newcommand{\anpi} {{\Lx\frac{\alpha_N}{\pi}\Rx}}
\newcommand{\ampi} {{\Lx\frac{\alpha_M}{\pi}\Rx}}
\newcommand{\aupi} {{\Lx\frac{\alpha_U}{\pi}\Rx}}
\newcommand{\betas}[1] {{\beta_{#1}}}

\newcommand{\betaN}[1] {{\beta^{N}_{#1}}}
\newcommand{\betaM}[1] {{\beta^{M}_{#1}}}
\newcommand{\betaU}[1] {{\beta^{U}_{#1}}}

\newcommand{\ket}[1]{\left| #1\right\rangle}

\newcommand{\CF}{C_F}
\newcommand{\CA}{C_A}
\newcommand{\nc}{N_c}
\newcommand{\cg}{{\cal G}}

\begin{document}

\today

\bibliographystyle{apsrev}

\title{The Two-Loop Infrared Structure of Amplitudes with Mixed Gauge Groups}

\author{William~B.~Kilgore}
\affiliation{
Physics Department, Brookhaven National Laboratory,
  Upton, New York
  11973, USA.\\
  {\tt [kilgore@bnl.gov]} }

\begin{abstract}
  The infrared structure of (multi-loop) scattering amplitudes is
  determined entirely by the identities of the external particles
  participating in the scattering.  The two-loop infrared structure of
  pure \qcd\ amplitudes has been known for some time.  By computing
  the two-loop amplitudes for $\overline{f}\,f\longrightarrow X$ and
  $\overline{f}\,f\longrightarrow V_1\,V_2$ scattering in an
  $SU(N)\times SU(M)\times U(1)$ gauge theory, I determine the
  anomalous dimensions which govern the infrared structure for any
  massless two-loop amplitude.
\end{abstract}

\maketitle

\section{Introduction}

The infrared structure of gauge theory amplitudes is governed by a set
of anomalous dimensions.  The anomalous dimensions at a particular
loop-level can be computed directly or extracted from a small number
of relatively simple amplitude calculations.  Once determined, these
anomalous dimensions allow one to predict, for any amplitude, no
matter how complex, the complete infrared structure to the given loop
level~\cite{Catani:1998bh,Sterman:2002qn}.  In \qcd, the anomalous
dimensions are known completely, in both the massless and massive
cases for one and two loop amplitudes, and their properties beyond the
two-loop level are being actively
studied~\cite{Aybat:2006wq,Aybat:2006mz,Mitov:2009sv,Becher:2009cu,
  Gardi:2009qi,Becher:2009qa,Becher:2009kw,Gardi:2009zv,Dixon:2009ur,
  Mitov:2010xw}.  Because of the many diagrams involved and the
complexity of the resulting amplitudes, foreknowledge of the infrared
structure is extremely valuable.  This knowledge was an important
guide for the ground-breaking calculations of two-loop parton
scattering
amplitudes~\cite{Bern:2000ie,Anastasiou:2000kg,Anastasiou:2000ue,
  Anastasiou:2001sv,Glover:2001af,Garland:2001tf,Anastasiou:2002zn,
  Glover:2003cm}.

Precision measurements in particle physics often involve the
interaction of more than one gauge group.  In particular, at hadron
colliders, nominally electroweak processes always involve some
interaction with \qcd.  Precision calculations of such processes,
therefore, require the computation of higher-order corrections in
mixed gauge groups~\cite{Kilgore:2011pa}.

In the current letter, I consider a theory with the following
structure: There are three gauge interactions, obeying an $SU(N)\times
SU(M)\times U(1)$ symmetry.  Fermions occur in four different
representations: $F_l$, which carry $U(1)$ charge $Q_l$ and are singlets
under $SU(N)$ and $SU(M)$; $F_n$, which are in the fundamental
representation of $SU(N)$, carry $U(1)$ charge $Q_n$ and are singlets
under $SU(M)$; $F_m$, which are in the fundamental representation of
$SU(M)$, carry $U(1)$ charge $Q_m$ and are singlets under $SU(N)$; and
$F_b$, which are in the fundamental representation of both $SU(N)$ and
$SU(M)$ and carry $U(1)$ charge $Q_b$.  Note that this is precisely
the structure of the (unbroken) Standard Model, where the $SU(N)$
theory corresponds to \qcd, the $SU(M)$ theory to the weak $SU(2)_L$
and the $U(1)$ to the hypercharge interaction.  Under this
identification, the $F_l$ multiplets correspond to the right-handed
leptons, the $F_m$ multiplets to the left-handed leptons, the $F_n$
multiplets to the right-handed quarks and the $F_b$ multiplets to the
left-handed quarks.

I will compute the two-loop amplitudes for $\overline{f}_x\,f_x
\longrightarrow X$ (where $X$ is a massive vector boson, neutral under
the $SU(N)\times SU(M)\times U(1)$ gauge symmetry) and
$\overline{f}_x\,f_x \longrightarrow V_1\,V_2$ for various
combinations of fermions and gauge bosons.  These calculations will
give me redundant extractions of the anomalous dimensions for each
particle type in the mixed gauge structure.  As a cross-check, I can
compare my results for the anomalous dimensions in a pure structure to
the known results in the literature.  All calculations are performed
in the conventional dimensional regularization
scheme~\cite{Collins:Renorm}.

\section{The infrared structure of \qcd\ amplitudes}
\label{sec:irstructure}
The infrared structure of pure \qcd\ interactions is well known.  For
a general $n$-parton scattering process, I label the set of external
partons by ${\bf f} = \{f_i\}_{i=1\dots n}$.  In the formulation of
Refs.~\cite{Sterman:2002qn,Aybat:2006wq,Aybat:2006mz}, a renormalized
amplitude may be factorized into three functions: the jet function
${\cal J}_{\bf f}$, which describes the collinear dynamics of the
external partons that participate in the collision; the soft function
${\bf S_f}$, which describes soft exchanges between the external
partons; and the hard-scattering function $\ket{H_{\bf f}}$, which
describes the short-distance scattering process,
\begin{equation}
\ket{{\cal M}_{\bf f}\Lx p_i,\tfrac{Q^2}{\mu^2},\alpha_s(\mu^2),\ep\Rx} =
   {\cal J_{\bf f}}\Lx\alpha_s(\mu^2),\ep\Rx\
    {\bf S_{f}}\Lx p_i,\tfrac{Q^2}{\mu^2},\alpha_s(\mu^2),\ep\Rx\
    \ket{H_{\bf f}\Lx p_i,\tfrac{Q^2}{\mu^2},\alpha_s(\mu^2)\Rx}\,.
\label{eqn:qcdfact}
\end{equation}
The notation indicates that $\ket{H_{\bf f}}$ is a vector and 
${\bf S_f}$ is a matrix in color space~\cite{Catani:1996jh,
Catani:1997vz,Catani:1998bh}.  As with any factorization, there is
considerable freedom to move terms about from one function to the
others.  It is convenient~\cite{Aybat:2006wq,Aybat:2006mz} to define
the jet and soft functions, ${\cal J}_{\bf f}$ and ${\bf S_f}$, so
that they contain all of the infrared poles but only contain infrared
poles, while all infrared finite terms, including those at
higher-order in $\ep$, are absorbed into $\ket{H_{\bf f}}$.

\subsection{The jet function in \qcd}
The jet function ${\cal J}_{\bf f}$ is found to be the product of
individual jet functions ${\cal J}_{f_i}$ for each of the external
partons,
\begin{equation}
{\cal J}_{\bf f}\Lx\alpha_s(\mu^2),\ep\Rx = \prod_{i\in{\bf{f}}}\
   {\cal J}_{i}\Lx\alpha_s(\mu^2),\ep\Rx\,.
\end{equation}
Each individual jet function is naturally defined in terms of the
anomalous dimensions of the Sudakov form factor~\cite{Sterman:2002qn},
\begin{equation}
\begin{split}
\ln{\cal J}_i\Lx\alpha_s(\mu^2),\ep\Rx &=
  -\aspi\LB\frac{1}{8\,\ep^2}\gamma_{K\,i}^{(1)} +
  \frac{1}{4\,\ep}\cg_i^{(1)}(\ep)\RB\\
  &\quad + \aspi^2\left\{
   \frac{\betas{0}}{8}\frac{1}{\ep^2}\LB\frac{3}{4\,\ep}\gamma_{K\,i}^{(1)}
    + \cg_i^{(1)}(\ep)\RB -
    \frac{1}{8}\LB\frac{\gamma_{K\,i}^{(2)}}{4\,\ep^2} + \frac{{\cal
	G}_i^{(2)}(\ep)}{\ep}\RB  \right\} + \dots\,,
\label{eqn:logjetqcd}
\end{split}
\end{equation}
where
\begin{equation}
\begin{split}
    \gamma_{K\,i}^{(1)} &= 2\,C_i,\quad
    \gamma_{K\,i}^{(2)} = C_i\,K = C_i\LB \CA\Lx\frac{67}{18} -
    \zeta_2\Rx - \frac{10}{9}T_f\,N_f\RB,\quad C_q \equiv \CF,\quad
    C_g \equiv \CA\,,\\
  \cg_{q}^{(1)} &= \frac{3}{2}\CF + \frac{\ep}{2}\CF\Lx8-\zeta_2\Rx, 
  \qquad \cg_{g}^{(1)} = 2\,\betas{0} - \frac{\ep}{2}\CA\,\zeta_2\,,\\
  \cg_{q}^{(2)} &= \CF^2\Lx\frac{3}{16} - \frac{3}{2}\zeta_2 +
    3\,\zeta_3\Rx + \CF\,\CA\Lx\frac{2545}{432} + \frac{11}{12}\zeta_2
    - \frac{13}{4}\zeta_3\Rx - \CF\,T_f\,N_f\Lx\frac{209}{108} +
    \frac{1}{3}\zeta_2\Rx\,,\\
  \cg_{g}^{(2)} &= 4\,\betas{1} + \CA^2\Lx\frac{10}{27} -
    \frac{11}{12}\zeta_2 - \frac{1}{4}\zeta_3\Rx +
    \CA\,T_f\,N_f\Lx\frac{13}{27} + \frac{1}{3}\zeta_2\Rx +
    \frac{1}{2}\CF\,T_f\,N_f\,,\\
  \betas{0} &= \frac{11}{12}C_A - \frac{1}{3}T_f\,N_f\,,\qquad
  \betas{1} = \frac{17}{24}C_A^2 - \frac{5}{12}C_A\,T_f\,N_f
           - \frac{1}{4}C_F\,T_f\,N_f
\label{eqn:qcdanomdims}
\end{split}
\end{equation}

Although $\cg_i$ and $\gamma_{K\,i}$ are defined through the
Sudakov form factor, they can be extracted from fixed-order
calculations~\cite{Gonsalves:1983nq,Kramer:1986sg,Matsuura:1987wt,
Matsuura:1989sm,Harlander:2000mg,Moch:2005id,Moch:2005tm}.
$\gamma_{K\,i}$ is the cusp anomalous dimension and represents a pure
pole term. The $\cg_i$ anomalous dimensions contain terms at
higher order in $\ep$, but I only keep terms in the expansion that
contribute poles to $\ln\Lx{\cal J}_i\Rx$.  $\beta_0$ and $\beta_1$
are the first two coefficients of the \qcd\ $\beta$-function,
$\CF = (\nc^2-1)/(2\,\nc)$ denotes the Casimir operator of the
fundamental representation of SU($\nc$), while $\CA=\nc$ denotes the
Casimir of the adjoint representation. $N_f$ is the number of quark
flavors and $T_f = 1/2$ is the normalization of the \qcd\ charge of
the fundamental representation.  $\zeta_n=\sum_{k=1}^{\infty}1/k^n$
represents the Riemann zeta-function of integer argument $n$.

\subsection{The soft function in \qcd}
The soft function is determined entirely by the soft anomalous
dimension matrix ${\bm\Gamma}_{S_f}$,
\begin{equation}
\begin{split}
{\bf S_f}\Lx p_i,\tfrac{Q^2}{\mu^2},\alpha_s(\mu^2),\ep\Rx
  &= 1 + \frac{1}{2\,\ep}\aspi{\bm\Gamma}_{S_f}^{(1)} +
   \frac{1}{8\,\ep^2}\aspi^2{\bm\Gamma}_{S_f}^{(1)}\times{\bm\Gamma}_{S_f}^{(1)}\\
  &\qquad- \frac{\betas{0}}{4\,\ep^2}\aspi^2{\bm\Gamma}_{S_f}^{(1)}
   + \frac{1}{4\,\ep}\aspi^2{\bm\Gamma}_{S_f}^{(2)} + \dots\,.
\label{eqn:qcdsoft}
\end{split}
\end{equation}
In the color-space notation of
Refs.~\cite{Catani:1996jh,Catani:1997vz,Catani:1998bh}, the soft
anomalous dimension is given by~\cite{Aybat:2006wq,Aybat:2006mz}
\begin{equation}
{\bm\Gamma}_{S_f}^{(1)} = \frac{1}{2}\,\sum_{i\in{\bf f}}\ \sum_{j\ne i}
   {\bf T}_i\cdot{\bf T}_j\,\ln\Lx\frac{\mu^2}{-s_{ij}}\Rx,\qquad
 {\bm\Gamma}_{S_f}^{(2)} = \frac{K}{2}{\bm\Gamma}_{S_f}^{(1)}\,,
\label{eqn:cdrsoftanomdim}
\end{equation}
where $K = \CA\Lx67/18-\zeta_2\Rx - 10\,T_f\,N_f/9$ is the same constant
that relates the one- and two-loop cusp anomalous dimensions. The 
${\bf T}_i$ are the color generators in the representation of parton $i$
(multiplied by $(-1)$ for incoming quarks and gluons and outgoing
anti-quarks).

\section{The infrared structure of mixed gauge groups}
When one includes additional gauge symmetries, the dominant effect on
the infrared structure is a replication of the \qcd\ structure, with
appropriate changes accounting for the size of the gauge group and the
Abelian character of the $U(1)$.  There are, however, new terms that
correspond to intrinsically mixed gauge interactions.  It is these
mixed terms I am interested in computing in this letter.  In
reference~\cite{Kilgore:2011pa}, some of the two-loop anomalous
dimensions for \qcd\,$\times$\qed\ amplitudes were determined, while
the forms of others, particularly those involving external gauge
bosons, were merely conjectured.  The current calculation explicitly
determines all of the two-loop mixed anomalous dimensions.

In a theory with the $SU(N)\times SU(M)\times U(1)$ symmetry described
above, the jet function for an external parton of species $i$ is
\begin{equation}
\begin{split}
\ln{\cal J}_i\Lx\alpha_N,\alpha_M,\alpha_U,\ep\Rx &=
  -\anpi\LB\frac{1}{8\,\ep^2}\gamma_{K\,i}^{(100)} +
  \frac{1}{4\,\ep}\cg_i^{(100)}(\ep)\RB\\
  &\quad + \anpi^2\left\{
   \frac{\betaN{200}}{8}\frac{1}{\ep^2}\LB\frac{3}{4\,\ep}\gamma_{K\,i}^{(100)}
    + \cg_i^{(100)}(\ep)\RB -
    \frac{1}{8}\LB\frac{\gamma_{K\,i}^{(200)}}{4\,\ep^2} + \frac{{\cal
	G}_i^{(200)}(\ep)}{\ep}\RB  \right\}\\
  &\quad -\ampi\LB\frac{1}{8\,\ep^2}\gamma_{K\,i}^{(010)} +
  \frac{1}{4\,\ep}\cg_i^{(010)}(\ep)\RB\\
  &\quad + \ampi^2\left\{
   \frac{\betaM{020}}{8}\frac{1}{\ep^2}\LB\frac{3}{4\,\ep}\gamma_{K\,i}^{(010)}
    + \cg_i^{(010)}(\ep)\RB -
    \frac{1}{8}\LB\frac{\gamma_{K\,i}^{(020)}}{4\,\ep^2} + \frac{{\cal
	G}_i^{(020)}(\ep)}{\ep}\RB  \right\}\\
  &\quad -\aupi\LB\frac{1}{8\,\ep^2}\gamma_{K\,i}^{(001)} +
  \frac{1}{4\,\ep}\cg_i^{(001)}(\ep)\RB\\
  &\quad + \aupi^2\left\{
   \frac{\betaU{002}}{8}\frac{1}{\ep^2}\LB\frac{3}{4\,\ep}\gamma_{K\,i}^{(001)}
    + \cg_i^{(001)}(\ep)\RB -
    \frac{1}{8}\LB\frac{\gamma_{K\,i}^{(002)}}{4\,\ep^2} + \frac{{\cal
	G}_i^{(002)}(\ep)}{\ep}\RB  \right\}\\
  &\quad - \anpi\,\ampi\,\LB\frac{1}{8\,\ep^2}\gamma_{K\,i}^{(110)}
               + \frac{1}{4\,\ep}\cg_i^{(110)}(\ep)\RB\\
  &\quad - \anpi\,\aupi\,\LB\frac{1}{8\,\ep^2}\gamma_{K\,i}^{(101)}
               + \frac{1}{4\,\ep}\cg_i^{(101)}(\ep)\RB\\
  &\quad - \ampi\,\aupi\,\LB\frac{1}{8\,\ep^2}\gamma_{K\,i}^{(011)}
               + \frac{1}{4\,\ep}\cg_i^{(011)}(\ep)\RB + \dots\,.
\label{eqn:logjetmnu}
\end{split}
\end{equation}
To deal with the multiplicity of gauge couplings, I have introduced
some new notations.  $\alpha_N$, $\alpha_M$, $\alpha_U$, are the
renormalized gauge couplings of the $SU(N)$, $SU(M)$ and $U(1)$
symmetries respectively.  Their $\beta$-function coefficients are
indexed by the powers of the gauge couplings (in $N,M,U$ order) that
multiply that coefficient.  For example,
\begin{equation}
\begin{split}
\beta^N(\alpha_N,\alpha_M,\alpha_U) &= \mu^2\frac{d}{d\mu^2}\anpi\\
   & = 
  - \anpi^2\betaN{200} - \anpi^3\betaN{300} - \anpi^2\ampi\betaN{210}
  - \anpi^2\aupi\betaN{201} + \dots\,,\\
\end{split}
\end{equation} 
where
\begin{equation}
\begin{split}
 \betaN{200} &= \frac{11}{12}C_{A_N} - \frac{1}{6}\Lx N_{f_n}+C_{A_M}\,N_{f_b}\Rx\,,\qquad
  \betaN{300} = \frac{17}{24}C_{A_N}^2 - \Lx\frac{5}{24}C_{A_N} + \frac{1}{8}C_{F_N}\Rx
   \Lx\,N_{f_n}+C_{A_M}\,N_{f_b}\Rx\,,\\
  \betaN{210} &= -\frac{1}{16}C_{A_M}\,N_{f_b}C_{F_M}\,,\qquad
  \betaN{201} = - \frac{1}{16}\Lx \sum_{i=1}^{N_{f_n}}Q_{f_n^i}^2
                + C_{A_M}\sum_{i=1}^{N_{f_b}}Q_{f_b^i}^2\Rx\,,
\end{split}
\end{equation} 
Similarly, the cusp ($\gamma_{K}$) and $\cg$ anomalous
dimensions are indexed by the powers of the gauge couplings that
multiply their leading appearance in the jet functions.  The explicit
values of all of the anomalous dimensions that appear through two
loops are given in Appendix~\ref{sec::IRAnomDims}.

The soft anomalous dimension of a mixed gauge structure, like the log
of the jet function, consists of the sum of the soft anomalous
dimensions for each of the separate gauge interactions, plus possible
terms that are due exclusively to the mixed interaction.  The
structure of such a mixed soft anomalous dimension would have to
involve (at least) pairs of generators from each of the mixing gauge
groups.  The least complicated of such terms would be of the form
\begin{equation}
\begin{split}
{\bm\Gamma}_{S_f}^{(110)} &= \frac{\digamma^{(110)}}{2}\,\sum_{i\in{\bf f}}\ \sum_{j\ne i}
   \Lx{\bf T_N}_i\cdot{\bf T_N}_j\Rx
   \Lx{\bf T_M}_i\cdot{\bf T_M}_j\Rx\,\ln\Lx\frac{\mu^2}{-s_{ij}}\Rx\\
{\bm\Gamma}_{S_f}^{(101)} &= \frac{\digamma^{(101)}}{2}\,\sum_{i\in{\bf f}}\ \sum_{j\ne i}
   \Lx{\bf T_N}_i\cdot{\bf T_N}_j\Rx
   Q_i\,Q_j\,\ln\Lx\frac{\mu^2}{-s_{ij}}\Rx
\end{split}
\end{equation}
The resulting soft function is
\begin{equation}
\begin{split}
{\bf S_f} = 1& + \anpi\frac{1}{2\,\ep}{\bm\Gamma}_{S_f}^{(100)}
     + \anpi^2\Lx\frac{1}{8\,\ep^2}{\bm\Gamma}_{S_f}^{(100)}
            \times{\bm\Gamma}_{S_f}^{(100)}
           - \frac{\betaN{200}}{4\,\ep^2}{\bm\Gamma}_{S_f}^{(100)}
           + \frac{1}{4\,\ep}{\bm\Gamma}_{S_f}^{(200)}\Rx\\
   & + \ampi\frac{1}{2\,\ep}{\bm\Gamma}_{S_f}^{(010)}
     + \ampi^2\Lx\frac{1}{8\,\ep^2}{\bm\Gamma}_{S_f}^{(010)}
            \times{\bm\Gamma}_{S_f}^{(010)}
           - \frac{\betaM{020}}{4\,\ep^2}{\bm\Gamma}_{S_f}^{(010)}
           + \frac{1}{4\,\ep}{\bm\Gamma}_{S_f}^{(020)}\Rx\\
   & + \aupi\frac{1}{2\,\ep}{\bm\Gamma}_{S_f}^{(001)}
     + \aupi^2\Lx\frac{1}{8\,\ep^2}{\bm\Gamma}_{S_f}^{(001)}
            \times{\bm\Gamma}_{S_f}^{(001)}
           - \frac{\betaU{002}}{4\,\ep^2}{\bm\Gamma}_{S_f}^{(001)}
           + \frac{1}{4\,\ep}{\bm\Gamma}_{S_f}^{(002)}\Rx\\
   & + \anpi\ampi\Lx\frac{1}{4\,\ep^2}{\bm\Gamma}_{S_f}^{(100)}
            \times {\bm\Gamma}_{S_f}^{(010)}
            + \frac{1}{4\,\ep}{\bm\Gamma}_{S_f}^{(110)}\Rx\\
   & + \anpi\aupi\Lx\frac{1}{4\,\ep^2}{\bm\Gamma}_{S_f}^{(100)}
            \times {\bm\Gamma}_{S_f}^{(001)}
            + \frac{1}{4\,\ep}{\bm\Gamma}_{S_f}^{(101)}\Rx\\
   & + \ampi\aupi\Lx\frac{1}{4\,\ep^2}{\bm\Gamma}_{S_f}^{(010)}
            \times {\bm\Gamma}_{S_f}^{(001)}
            + \frac{1}{4\,\ep}{\bm\Gamma}_{S_f}^{(011)}\Rx
\end{split}
\end{equation}
Any new terms that might arise from mixing are parameterized by
${\bm\Gamma}_{S_f}^{(110)}$, ${\bm\Gamma}_{S_f}^{(101)}$ and
${\bm\Gamma}_{S_f}^{(011)}$.

\section{Extracting the anomalous dimensions}
I will extract the anomalous dimensions be performing a few,
relatively simple, explicit calculations.  The anomalous dimensions
associated with the fermions can be extracted from a Sudakov-type
calculation, $\overline{f}_x\,f_x\longrightarrow X$, where $X$ is a
massive vector boson that is uncharged under the $SU(N)\times
SU(M)\times U(1)$ symmetry.  In this case the infrared structure of
the amplitude is uniquely associated with the $f_x$ fermions.
Alternatively, one could extract the fermion anomalous dimensions from
a set of calculations of the form $\overline{f}_l\,f_l \longrightarrow
\overline{f}_x\,f_x$.  For instance, because $f_l$ carries only the
$U(1)$ charge, the mixed infrared structure can again be uniquely
associated with the $f_x$ fermions.  This is the method used in
Ref.~\cite{Kilgore:2011pa}, where the $SU(3)\times U(1)$ anomalous
dimensions were determined from the mixed corrections to
$\overline{q}q\longrightarrow l^+l^-$.

I could extract the boson anomalous dimensions from another
Sudakov-type calculation, that of ``Higgs'' production,
$V_i\,V_i\longrightarrow H$.  The problem with this calculation is
that the scalar must either carry quantum numbers of the vector boson,
in which case it contributes to the infrared structure of the
amplitude, or it must couple to the vectors through an effective
interaction, for which one would need to determine the renormalization
properties and Wilson coefficients.  I will instead extract the gauge
boson anomalous dimensions from calculations of the more complicated
amplitudes, $\overline{f}_x\,f_x\longrightarrow V_1\,V_2$.  The
extraction of the boson anomalous dimensions from these amplitudes is
made simpler by the fact that I have already determined the fermion
anomalous dimensions from Sudakov-type amplitudes.

\subsection{Extracting the fermion anomalous dimensions}
The fermion anomalous dimensions are extracted from calculations of
the Sudakov-type amplitudes $\overline{f}_x\,f_x\longrightarrow X$.
\begin{figure}[h]
\includegraphics[height=3.0cm]{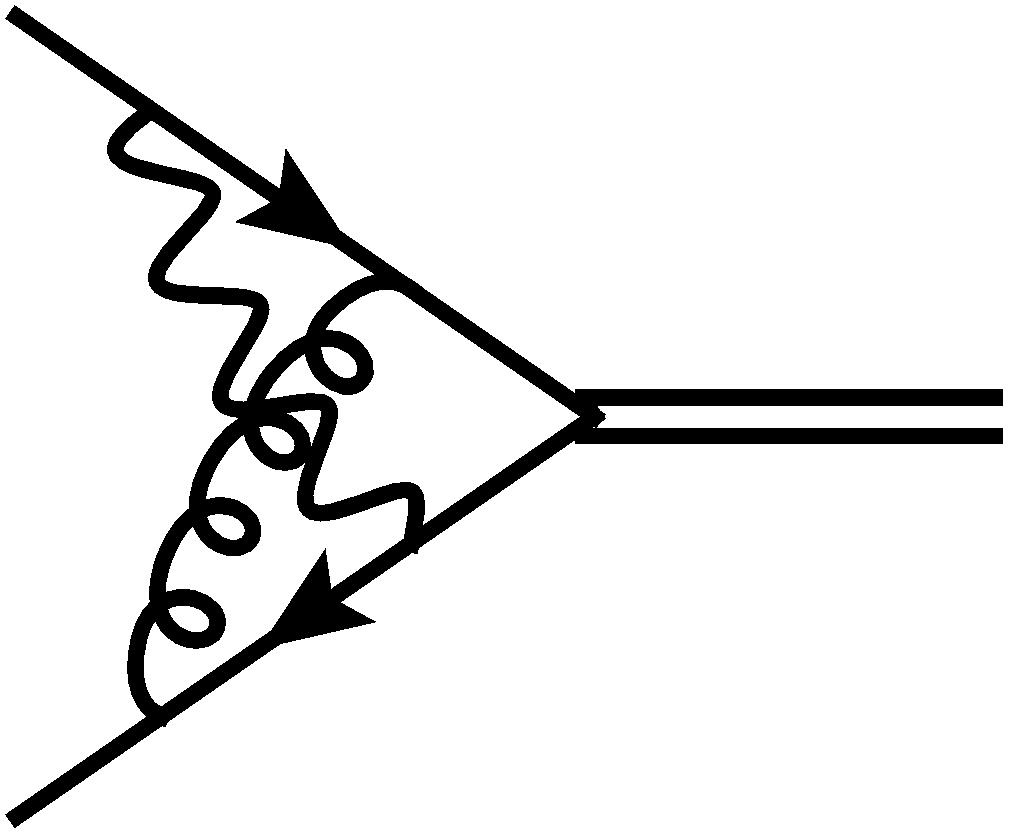}\qquad\qquad
\includegraphics[height=3.0cm]{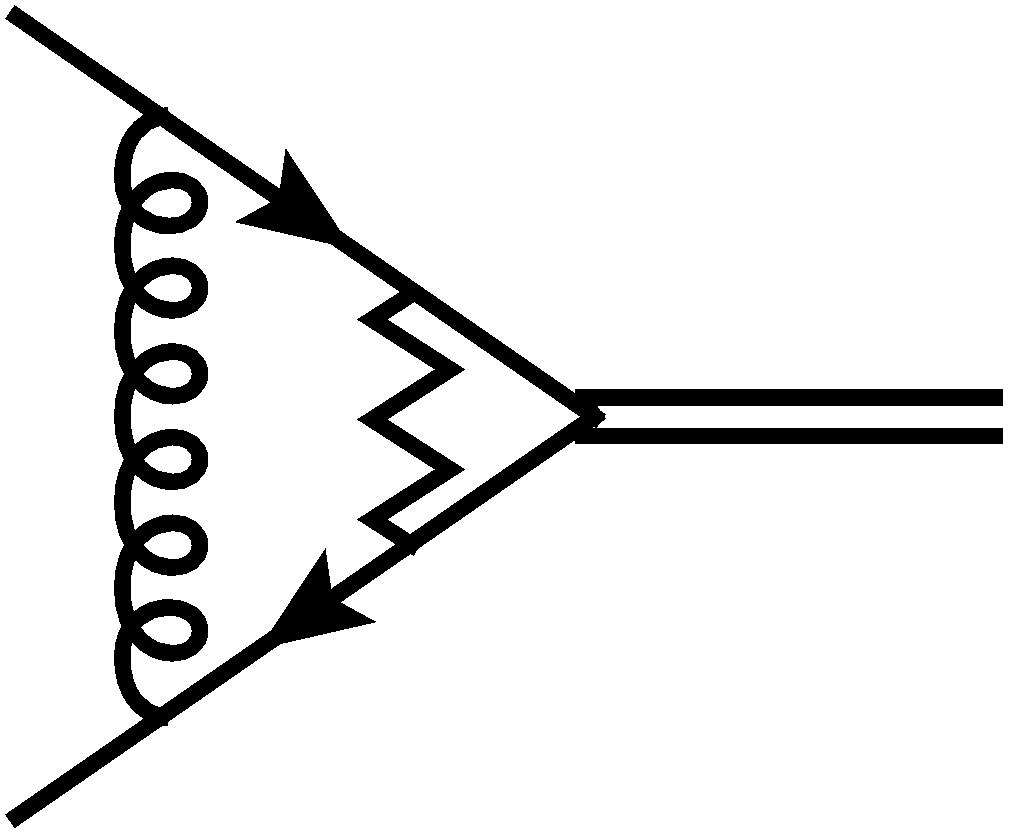}\qquad\qquad
\includegraphics[height=3.0cm]{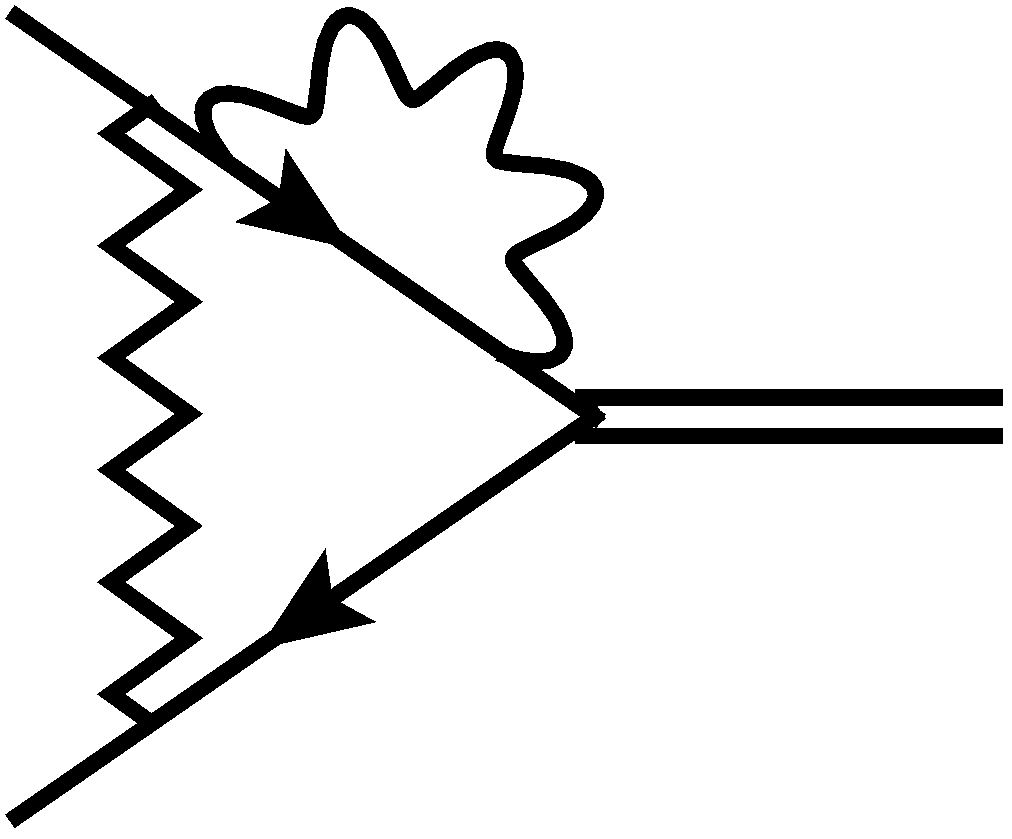}
\caption{Sample diagrams of $\overline{f}_b\,f_b\longrightarrow X$
\label{fig::FFVamps}}
\end{figure}
The Feynman diagrams (see Fig.~\ref{fig::FFVamps}) are essentially
the same as for two-loop \qcd\ corrections to Drell-Yan production.  I
generate the Feynman diagrams using QGRAF~\cite{Nogueira:1991ex} and
implement the Feynman rules and perform algebraic manipulations with
FORM~\cite{Vermaseren:2000nd}.  The resulting loop integrals are
reduced to master integrals using the integration-by-parts (IBP)
method~\cite{Chetyrkin:1981qh} in combination with Laporta's
algorithm~\cite{Laporta:1996mq,Laporta:2001dd} as implemented in the
program REDUZE2~\cite{vonManteuffel:2012np}.

\begin{figure}[h]
\includegraphics[height=2.75cm]{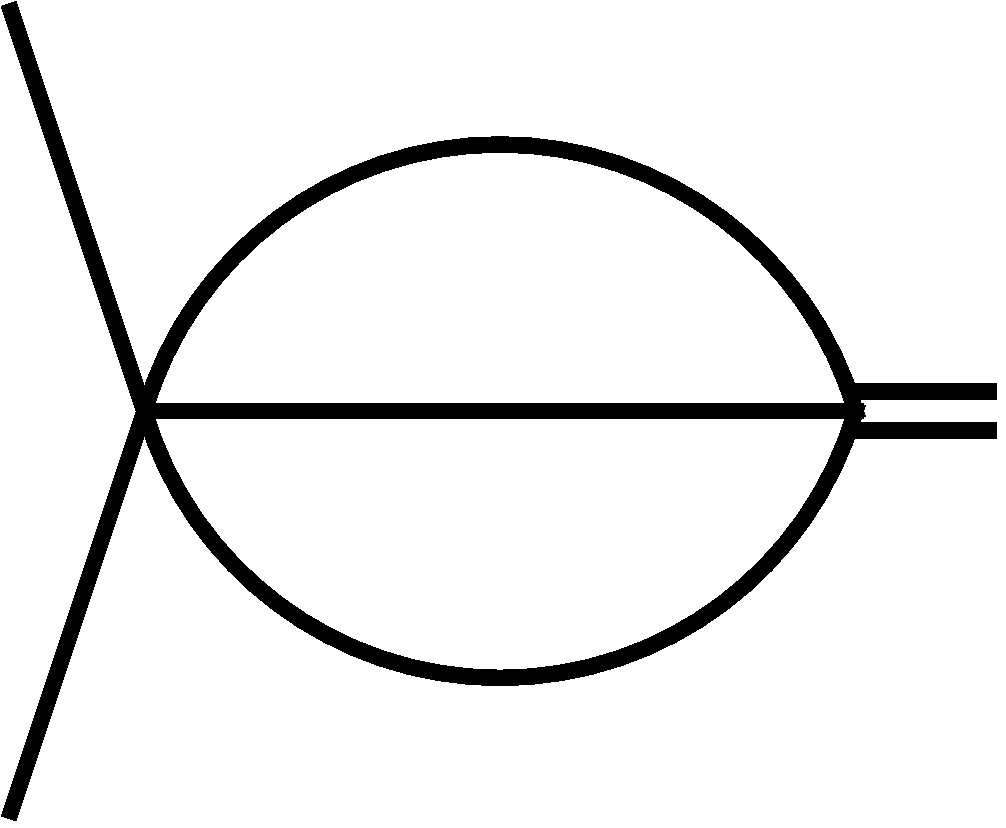}\qquad
\includegraphics[height=2.75cm]{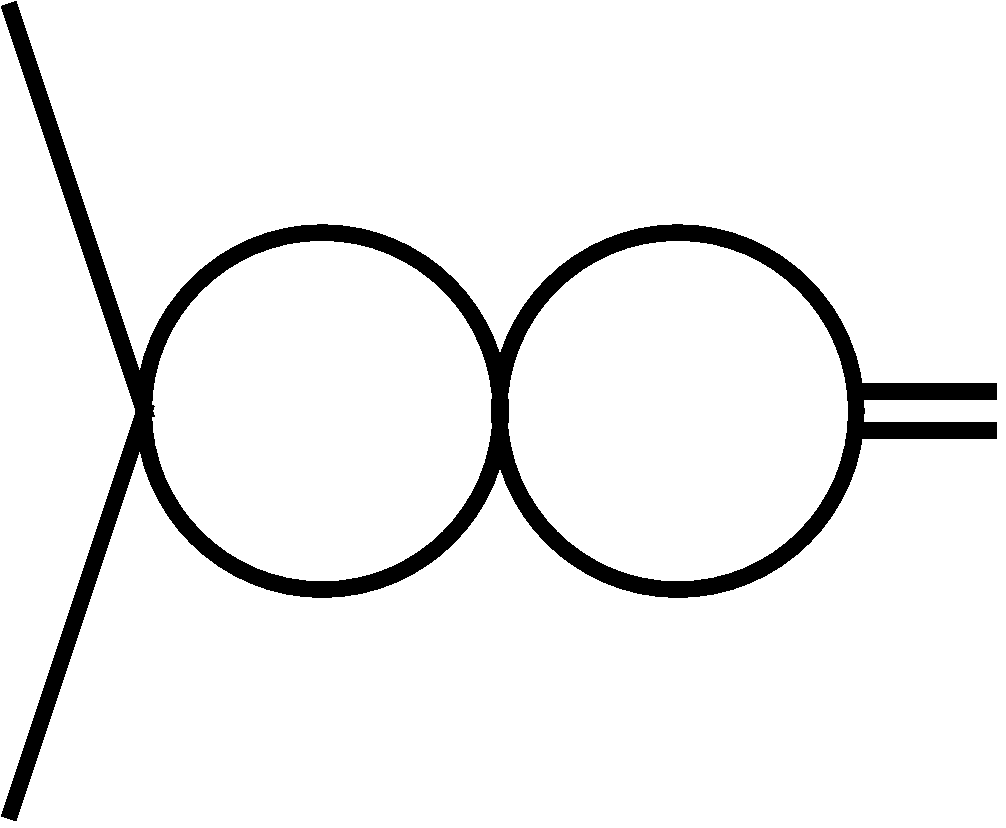}\qquad
\includegraphics[height=2.75cm]{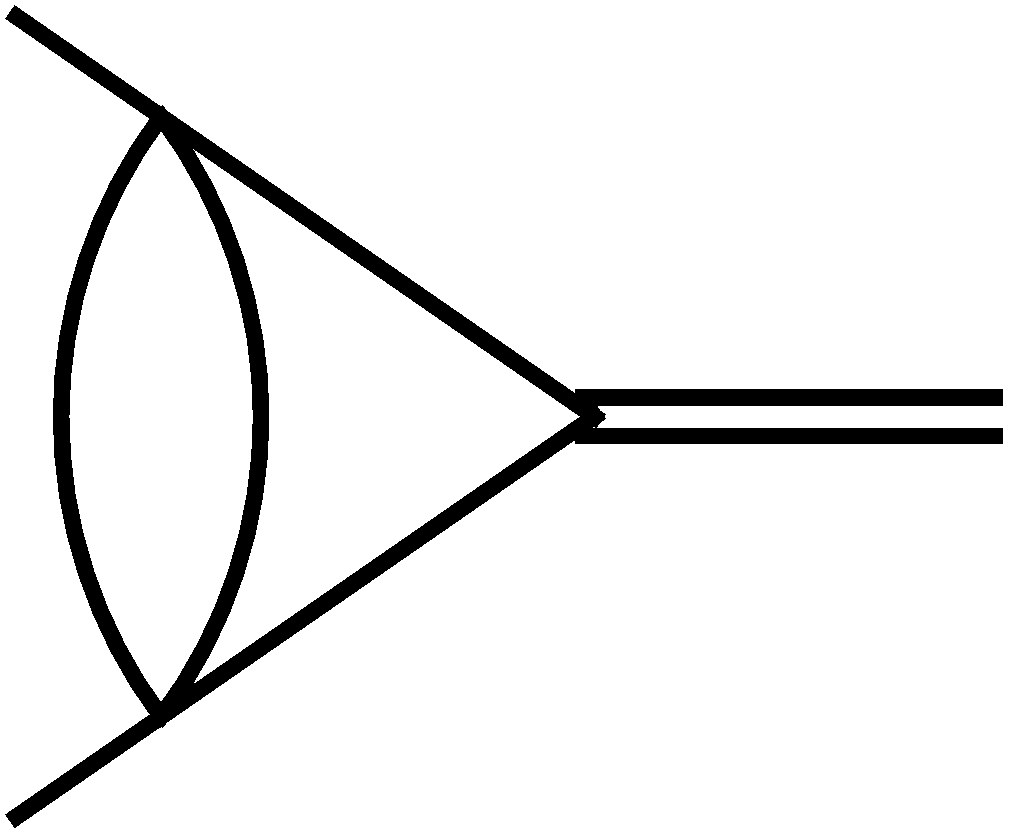}\qquad
\includegraphics[height=2.75cm]{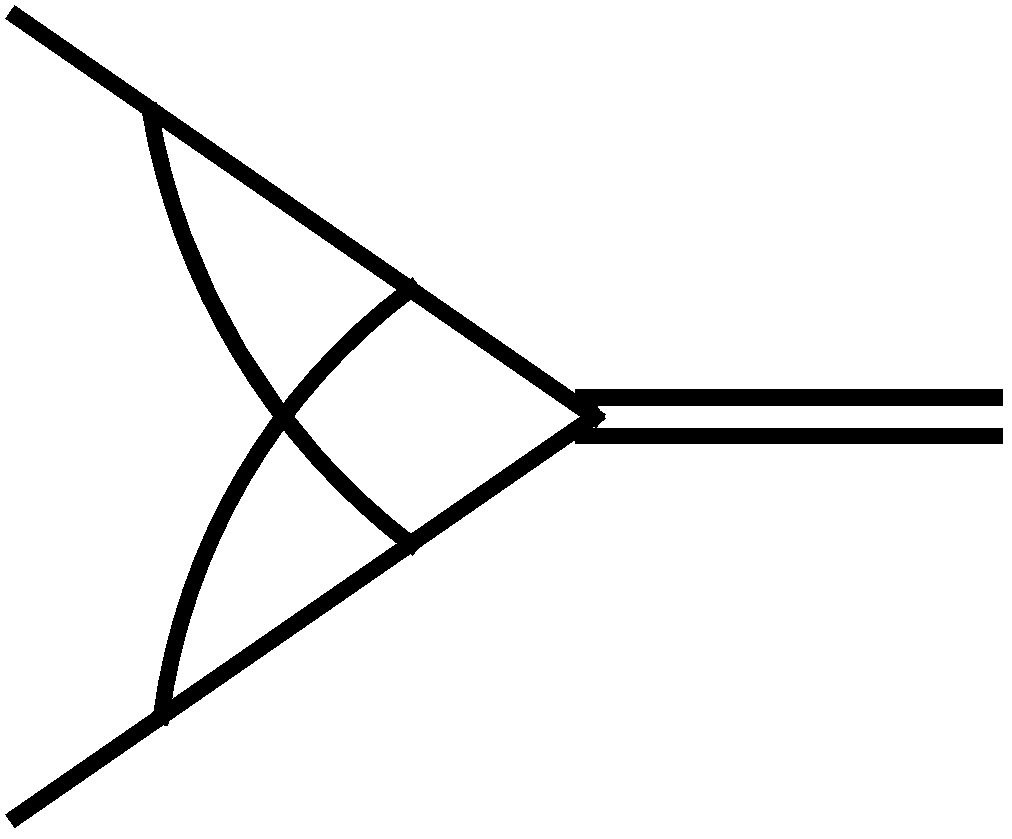}
\caption{Master Integrals for two-loop Sudakov-type amplitudes.
\label{fig::2to1masters}}
\end{figure}

There are only four master integrals (see
Fig.~\ref{fig::2to1masters}) that contribute to these
processes and all can be evaluated in closed form by standard Feynman
parameter integrals.  The results of the reduction to master integrals
and the values of the master integrals are inserted into the FORM
program, and the amplitude is evaluated as a Laurent series in the
dimensional regularization parameter $\ep$.  After renormalization,
the poles in $\ep$ are entirely infrared in origin.  Most of the
infrared terms can be readily associated with pure $SU(N)$, $SU(M)$ or
$U(1)$ interactions, or with the overlap of two one-loop terms.  Once
these terms are accounted for, however, one obtains the two-loop
mixed contribution to the fermion anomalous dimensions.  I find that
there are no mixed cusp anomalous dimensions for the fermions, nor is
there a mixed soft anomalous dimension involving only fermions.  There
are, however, mixed $\cg$ anomalous dimensions.  The results are
collected in Appendix~\ref{sec::IRAnomDims}.

\subsection{Extracting the boson anomalous dimensions}
\begin{figure}[h]
\includegraphics[height=2.50cm]{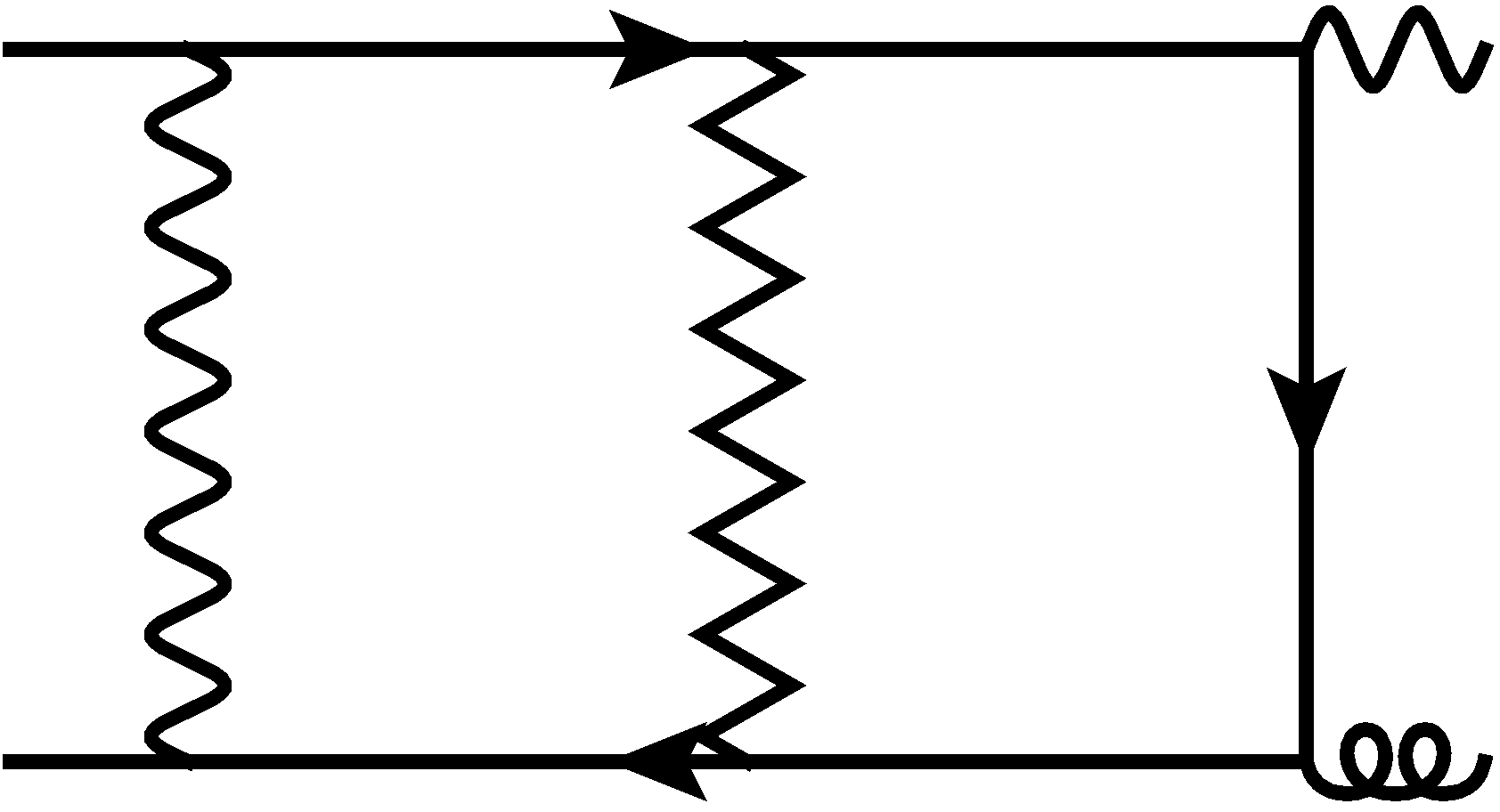}\qquad
\includegraphics[height=2.50cm]{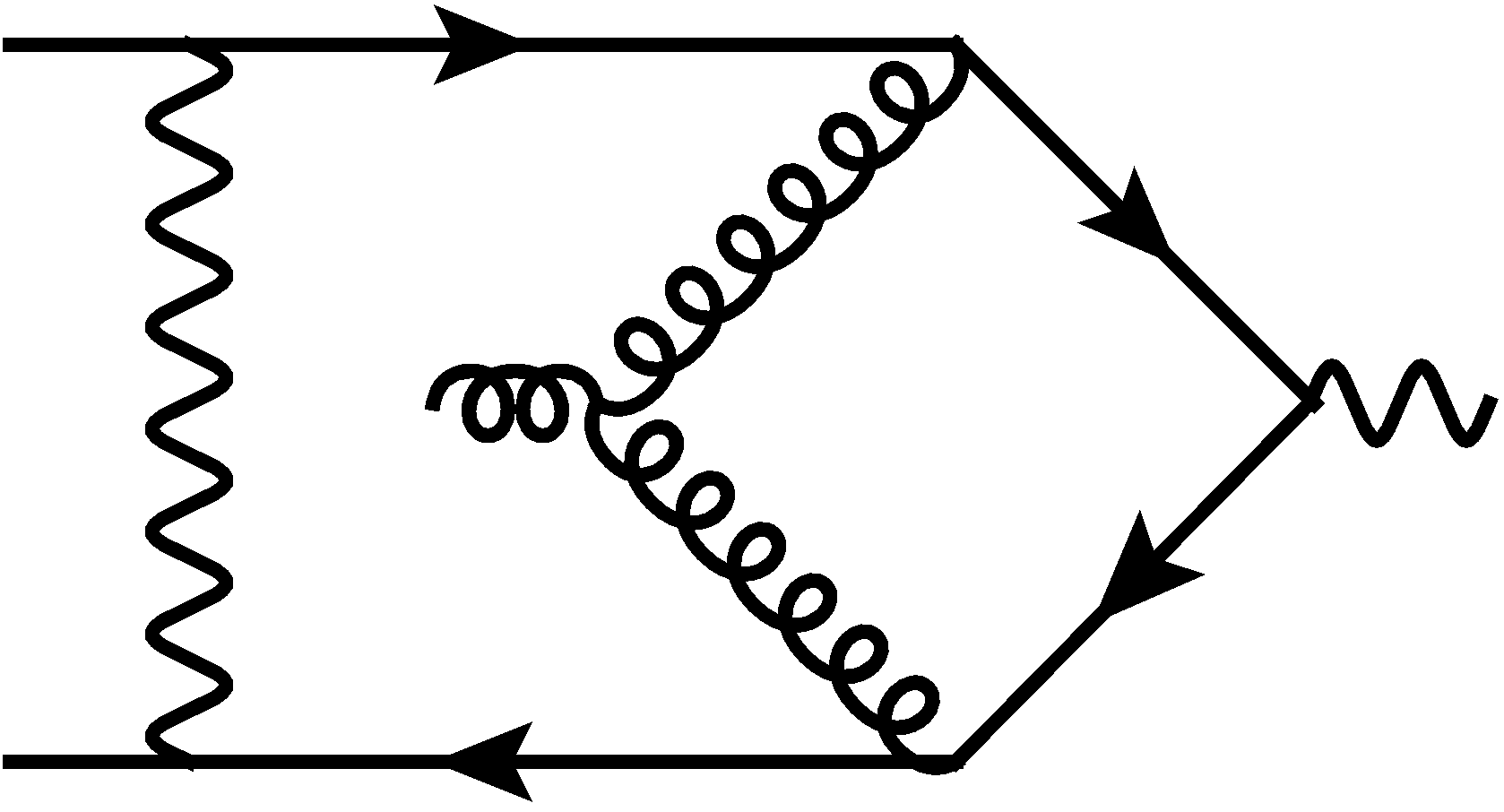}\qquad
\includegraphics[height=2.50cm]{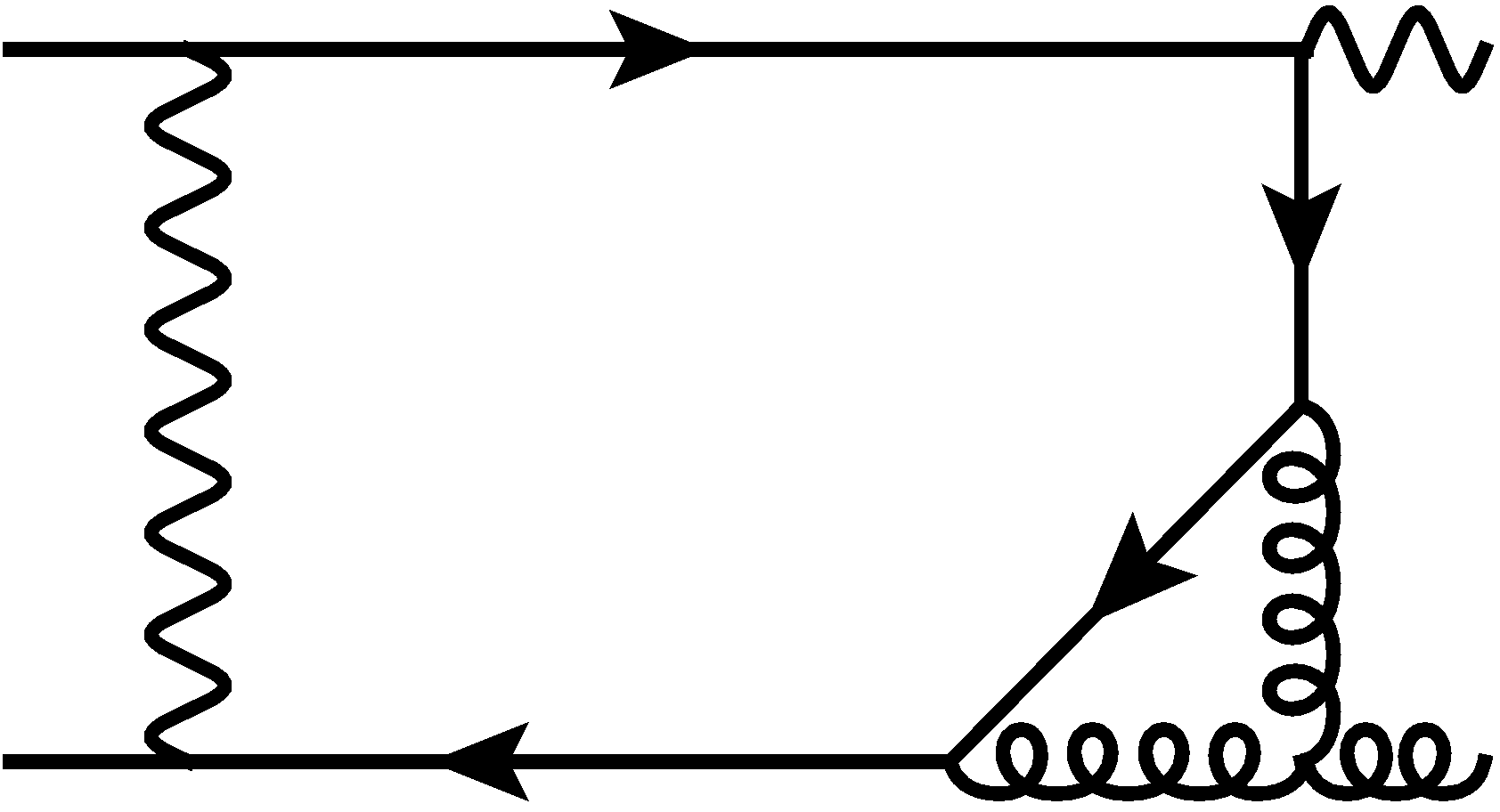}
\caption{Sample diagrams of $\overline{f}_b\,f_b\longrightarrow A_N\,A_M$
\label{fig::FFVVamps}}
\end{figure}

\begin{figure}[h]
\includegraphics[height=2.5cm]{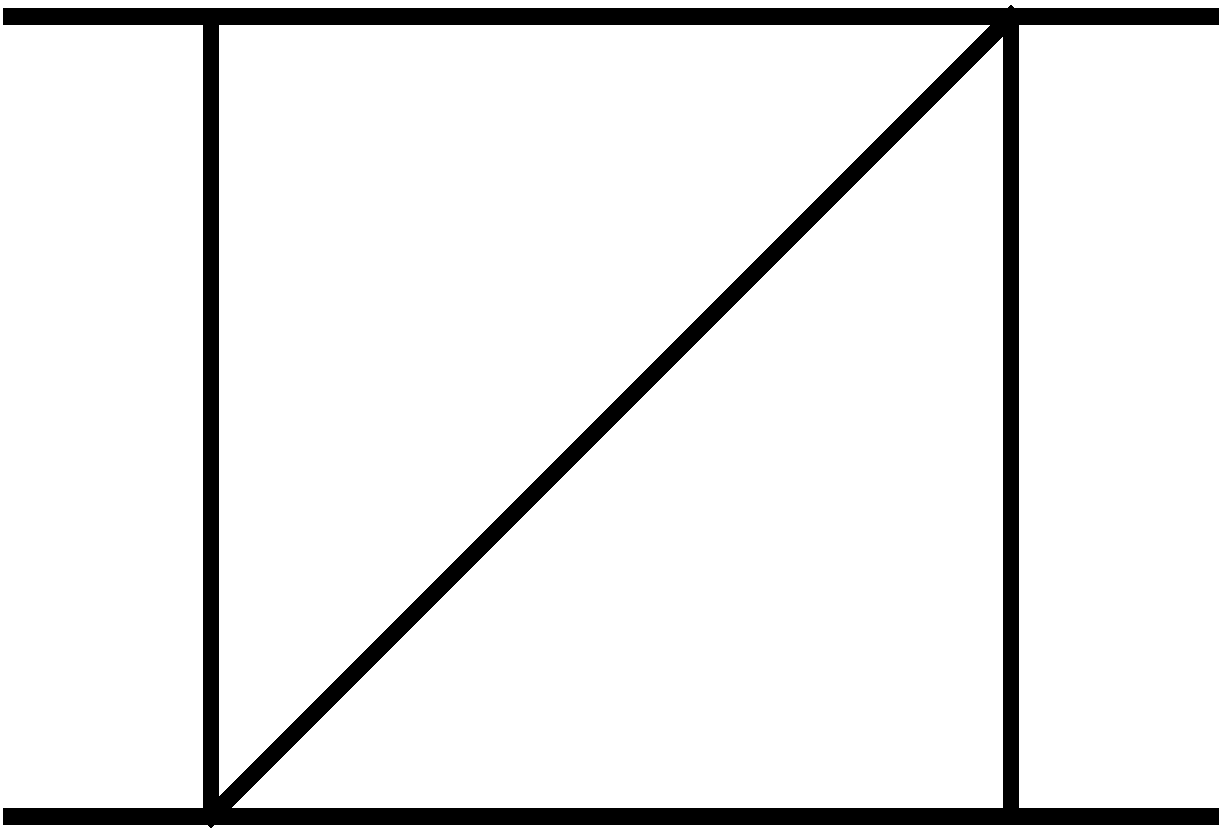}\qquad
\includegraphics[height=2.5cm]{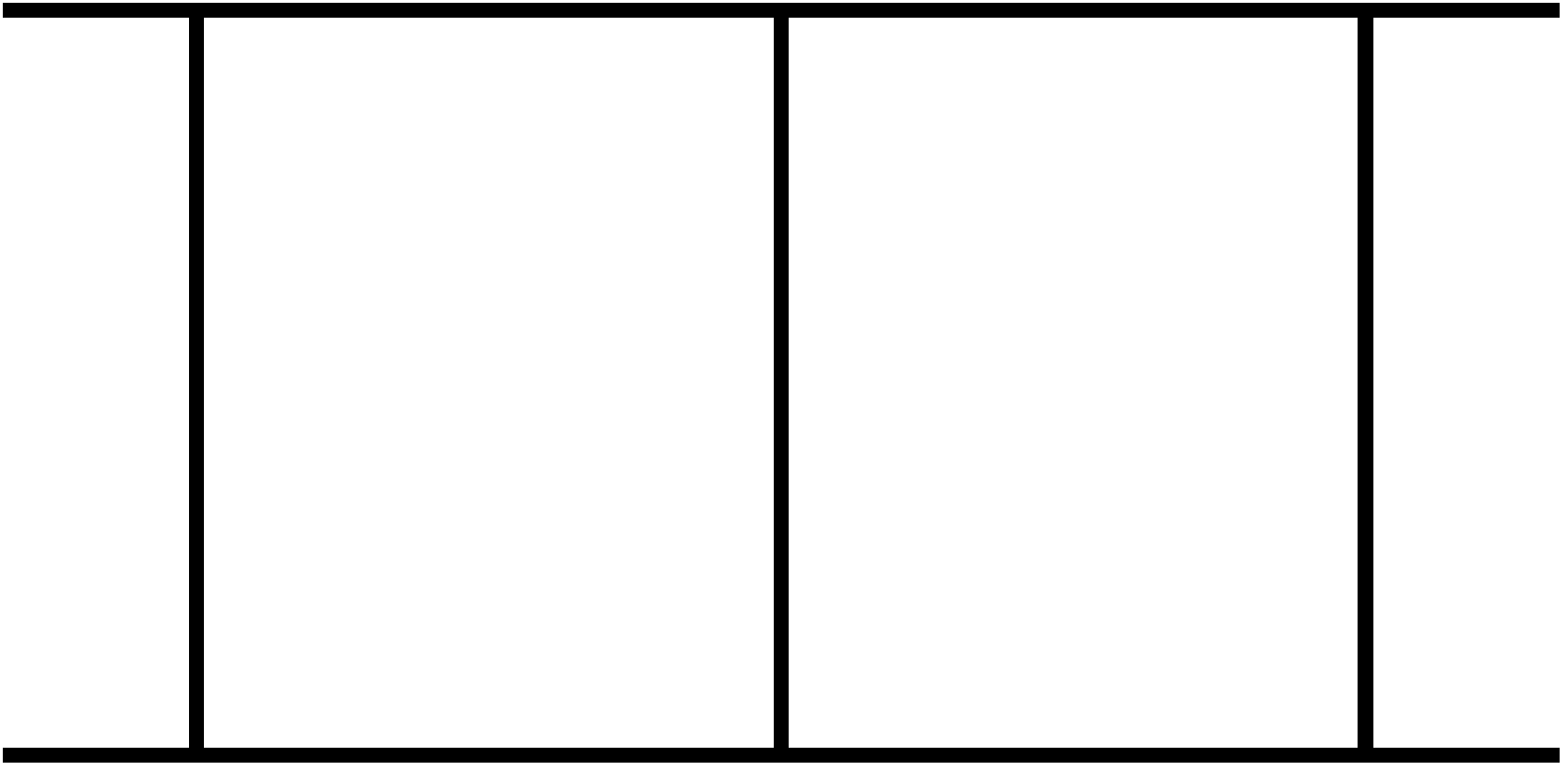}\qquad
\includegraphics[height=2.5cm]{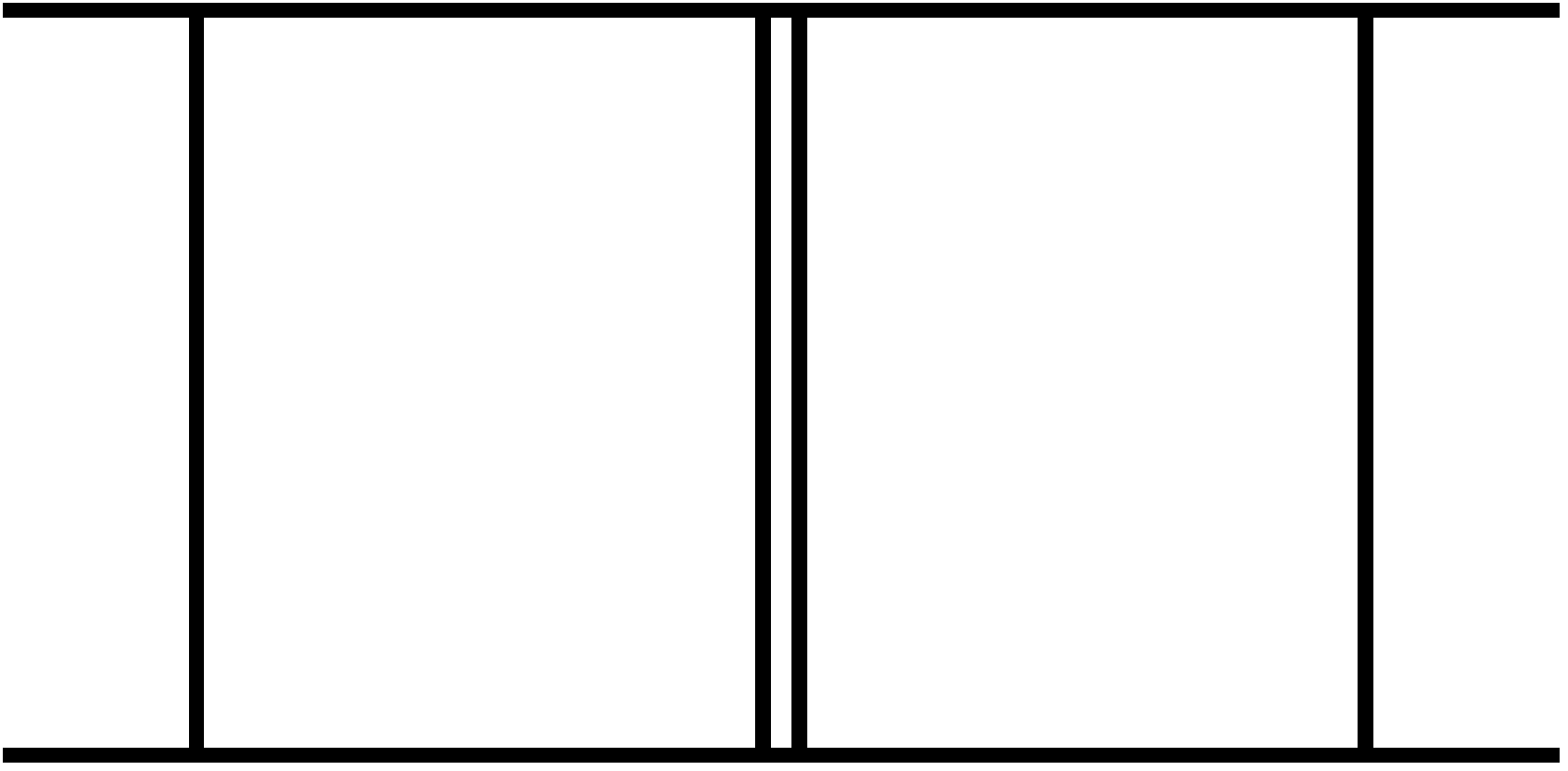}\\[10pt]
\includegraphics[height=2.5cm]{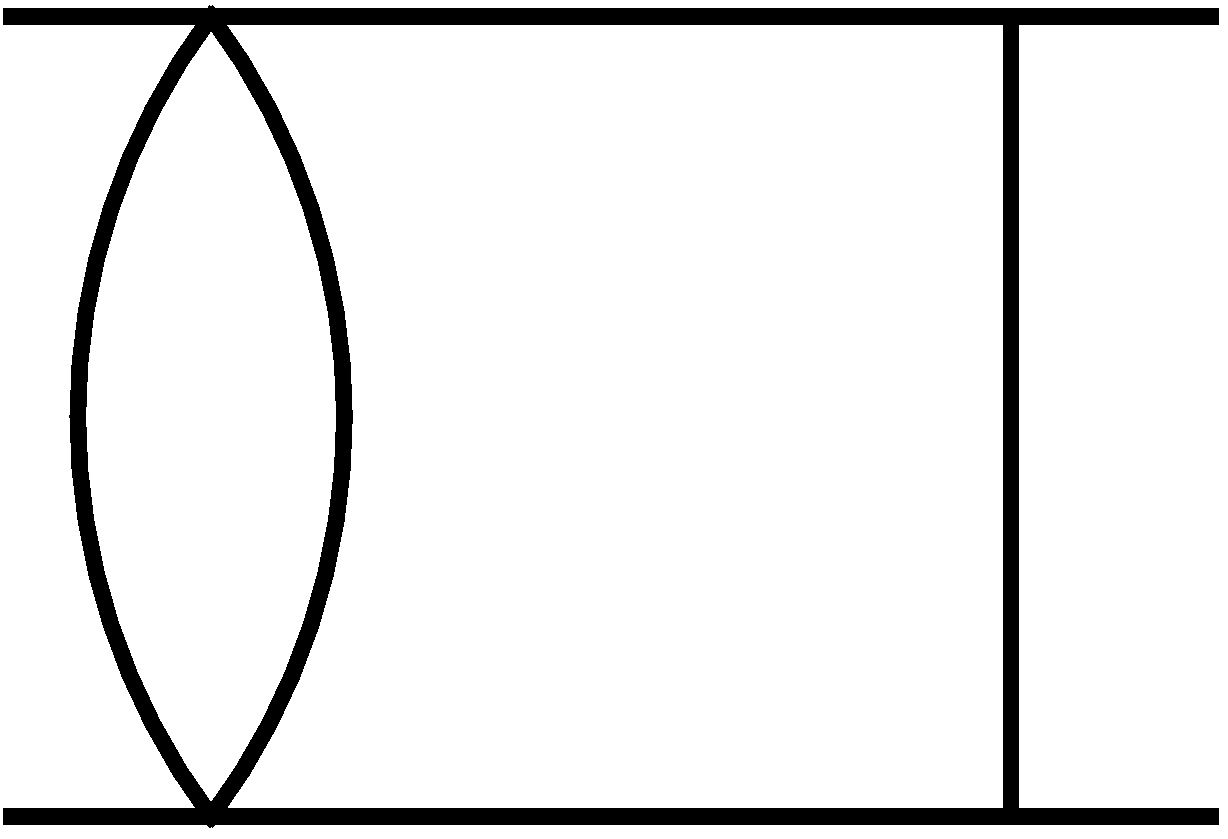}\qquad
\includegraphics[height=2.5cm]{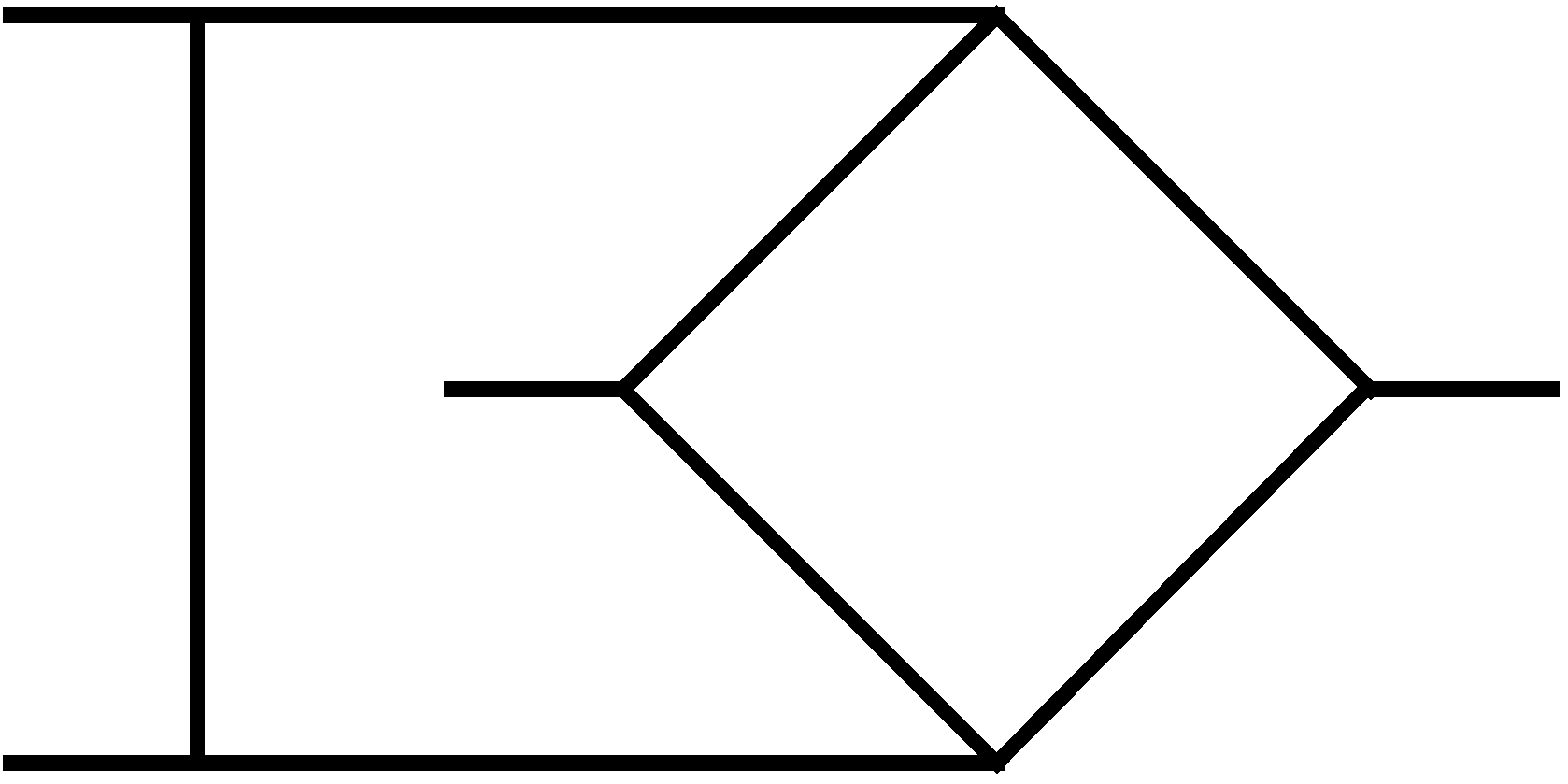}\qquad
\includegraphics[height=2.5cm]{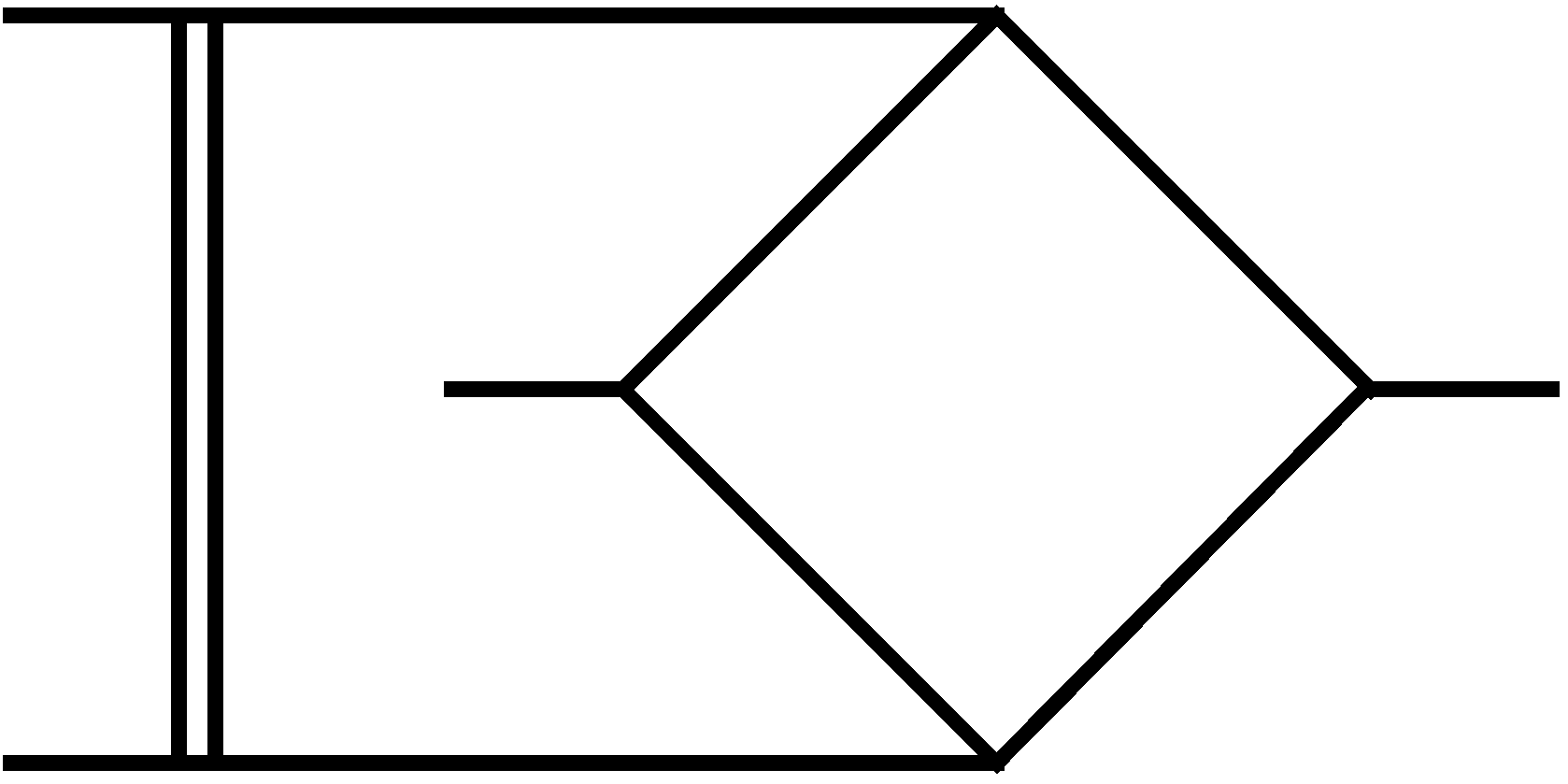}
\caption{Master integrals for two-loop massless two-to-two
  scattering.  The double lines indicate a squared propagator.
\label{fig::2to2masters}}
\end{figure}

The boson anomalous dimensions are extracted from two-loop, two-to-two
fermion to di-boson scattering amplitudes.  Sample diagrams are shown
in Fig.~\ref{fig::FFVVamps}.
In addition to the four master integrals that contribute to two-loop
Sudakov-type diagrams, there are six more that contribute to massless
two-to-two scattering (see Fig.~\ref{fig::2to2masters}).
In this case the infrared structure of the amplitudes involves the
overlap of the infrared structure of the fermions and the two gauge
bosons.  The soft anomalous dimensions can be identified by their
dependence on the logs of kinematic invariants.  The gauge boson
contributions to the jet functions must be determined by taking
different combinations of the external gauge bosons and accounting for
the contributions of the already-determined quark anomalous
dimensions.  As with the quarks, I find that there are no mixed cusp
or soft anomalous dimensions at two loops, but that there are
non-vanishing mixed $\cg$ anomalous dimensions.

\section{Conclusion}
I have computed the anomalous dimensions that govern the two-loop
infrared structure of mixed gauge interactions.  I have presented
results for a general $SU(N)\times SU(M)\times U(1)$ gauge structure
with fermions that lie in the fundamental representations of both
non-Abelian gauge groups ($F_b$), the fundamental representation of
one and the singlet representation of the other ($F_n$ and $F_m$), or
are singlets under both non-Abelian gauge groups ($F_l$).  All
fermions are assumed to carry $U(1)$ charges.  I note that this is the
gauge structure and fermion content of the unbroken Standard Model.
However, I have treated the fermions as vector-like, and therefore do
not have the chiral structure of the Standard Model.  Since the chiral
anomaly and anomaly cancellation are ultraviolet issues, they should
not affect the infrared structure at all.  If one were to make the
fermion multiplets chiral, so that $F_b$ and $F_m$ represent the
left-handed quarks and leptons, respectively, while $F_n$ and $F_l$
represent the right-handed quarks and leptons, one would only need to
weight factors of $N_{f_x}$ by a factor of $1/2$ to account for the
chiral projector in the fermion trace.  Since I have expressed the
anomalous dimensions so that explicit factors of $N_{f_x}$ only appear
in the coefficients of the $\beta$-functions, it is only there that
one would need to make this change.  The rest of the formul\ae\ in
Appendix~\ref{sec::IRAnomDims} remain unchanged.

The connection of the current results to applications in \qcd$\times$
\qed\ is more direct.  Here, I can identify the $SU(N)$ symmetry as
\qcd, and the $U(1)$ as \qed\ and drop the $SU(M)$ interaction.  In
this case, I need only $F_n$ and $F_l$ vector-like representations of
fermions.  One can readily check that the mixed $\cg$ anomalous
dimensions determined here agree with those determined for the quarks
in Reference~\cite{Kilgore:2011pa}.

The results determined here are not surprising and were largely
anticipated in Reference~\cite{Kilgore:2011pa} by examining the
structure of the \qcd\ anomalous dimensions.  The argument was that
there can be no non-Abelian structure in the mixed terms because the
generators of the different gauge groups commute with one another and
two-loop amplitudes are not sufficiently complicated to allow both
mixed interactions and non-Abelian structures of a single gauge group
in the same term.  Therefore, all factors of $C_A$ that appear in the
two-loop \qcd\ anomalous dimensions should be set to zero.
Furthermore, it was postulated that all factors of $N_f$ that appear
should be associated with coefficients of the $\beta$-functions.
However, contributions to two-loop anomalous dimensions that might
arise from corrections to one-loop terms would only involve leading
coefficients of the $\beta$-functions.  Because of the Ward identity,
mixing first appears in the $\beta$-functions of gauge couplings at
second order.  Therefore, corrections that are proportional to leading
coefficients of the $\beta$-functions should also be set to zero.

From this, one expects that there will be no mixed cusp or soft
anomalous dimensions at two-loops.  The factor $K$ which governs the
two-loop corrections to both of these terms can be written as a linear
combination of $C_A$ and the leading coefficient of the
$\beta$-function.  Thus, by this reasoning, the only mixed anomalous
dimensions that one expects at two-loops are $\cg$ terms.  If I assume
that the mixed $\cg$ anomalous dimensions will have essentially the
same form as those of \qcd, the only terms that remain are
proportional to $C_F^2$ or to $\beta_1$.  The minimal possible change
that is consistent with the mixed terms is to change each factor of
$C_F$ to one of $\{C_{F_N},C_{F_M},Q_f^2\}$ and to change $\beta_1$ to
the appropriate one of $\{\betaN{110},\betaN{101},\betaM{110},
\betaM{011},\betaU{101},\betaU{011}\}$.  It turns out that these
simple transformations give exactly the correct result.

\paragraph*{Acknowledgments:}
This research was supported by the U.S.~Department of Energy under
Contract No.~DE-AC02-98CH10886.

\appendix
\section{Infrared Anomalous Dimensions}
\label{sec::IRAnomDims}
\subsection{$\beta$-Functions}
The $\beta$-function of the $SU(N)$ coupling is
\begin{equation}
\begin{split}
\beta^N(\alpha_N,\alpha_M,\alpha_U) &= \mu^2\frac{d}{d\mu^2}\anpi\\
   & = 
  - \anpi^2\betaN{200} - \anpi^3\betaN{300} - \anpi^2\ampi\betaN{210}
  - \anpi^2\aupi\betaN{201} + \dots\,,\\
\end{split}
\end{equation} 
where
\begin{equation}
\begin{split}
  \betaN{200} &= \frac{11}{12}C_{A_N}
       - \frac{1}{3}\,T_f\Lx N_{f_n}+C_{A_M}\,N_{f_b}\Rx\,,\qquad
  \betaN{300} = \frac{17}{24} C_{A_N}^2 - \Lx\frac{5}{12}C_{A_N}
       + \frac{1}{4}C_{F_N}\Rx T_f
   \Lx\,N_{f_n}+C_{A_M}\,N_{f_b}\Rx\,,\\
  \betaN{210} &= -\frac{1}{8}C_{F_M}\,C_{A_M}\,T_f\,N_{f_b}\,,\qquad
  \betaN{201} = - \frac{1}{16}\Lx \sum_{i=1}^{N_{f_n}}Q_{f_n^i}^2
                + C_{A_M}\sum_{i=1}^{N_{f_b}}Q_{f_b^i}^2\Rx\,,
\end{split}
\end{equation}
For the $SU(M)$ coupling,
\begin{equation}
\begin{split}
\beta^M(\alpha_N,\alpha_M,\alpha_U) &= \mu^2\frac{d}{d\mu^2}\ampi\\
   & = 
  - \ampi^2\betaM{020} - \ampi^3\betaM{030} - \anpi\ampi^2\betaM{120}
  - \ampi^2\aupi\betaM{021} + \dots\,,\\
\end{split}
\end{equation} 
where
\begin{equation}
\begin{split}
  \betaM{020} &= \frac{11}{12}C_{A_M}
       - \frac{1}{3}\,T_f\Lx N_{f_m}+C_{A_N}\,N_{f_b}\Rx\,,\qquad
  \betaM{030} = \frac{17}{24} C_{A_M}^2 - \Lx\frac{5}{12}C_{A_M}
       + \frac{1}{4}C_{F_M}\Rx T_f
   \Lx\,N_{f_m}+C_{A_N}\,N_{f_b}\Rx\,,\\
  \betaM{120} &= -\frac{1}{8}C_{F_N}\,C_{A_N}\,T_f\,N_{f_b}\,,\qquad 
  \betaM{021} = - \frac{1}{16}\Lx \sum_{i=1}^{N_{f_m}}Q_{f_m^i}^2
                + C_{A_N}\sum_{i=1}^{N_{f_b}}Q_{f_b^i}^2\Rx\,,
\end{split}
\end{equation}
while for the $U(1)$,
\begin{equation}
\begin{split}
\beta^U(\alpha_N,\alpha_M,\alpha_U) &= \mu^2\frac{d}{d\mu^2}\ampi\\
   & = 
  - \ampi^2\betaU{020} - \ampi^3\betaU{030} - \anpi\ampi^2\betaU{120}
  - \ampi^2\aupi\betaU{021} + \dots\,,\\
\end{split}
\end{equation} 
where
\begin{equation}
\begin{split}
  \betaU{020} &= - \frac{1}{3}\Lx \sum_{i=1}^{N_{f_l}}Q_{f_l^i}^2
     + C_{A_M}\sum_{i=1}^{N_{f_m}}Q_{f_m^i}^2
     + C_{A_N}\sum_{i=1}^{N_{f_n}}Q_{f_n^i}^2
     + C_{A_N}C_{A_M}\sum_{i=1}^{N_{f_b}}Q_{f_b^i}^2\Rx\,,\\
  \betaU{030} &= - \frac{1}{4}\Lx \sum_{i=1}^{N_{f_l}}Q_{f_l^i}^4
     + C_{A_M}\sum_{i=1}^{N_{f_m}}Q_{f_m^i}^4
     + C_{A_N}\sum_{i=1}^{N_{f_n}}Q_{f_n^i}^4
     + C_{A_N}C_{A_M}\sum_{i=1}^{N_{f_b}}Q_{f_b^i}^4\Rx\,,\\
  \betaU{102} &= - \frac{1}{8}C_{A_N}\Lx \sum_{i=1}^{N_{f_n}}\,Q^2_{f_n^i}
     + C_{A_M}\,\sum_{i=1}^{N_{f_b}}\,Q^2_{f_b^i}\Rx \,,\qquad
  \betaU{012} = - \frac{1}{8}C_{A_M}\Lx \sum_{i=1}^{N_{f_m}}\,Q^2_{f_m^i}
     + C_{A_N}\,\sum_{i=1}^{N_{f_b}}\,Q^2_{f_b^i}\Rx \,.
\end{split}
\end{equation}
\subsection{The Cusp Anomalous Dimensions}
\begin{equation}
\begin{split}
  \gamma_{K\,f_n}^{(100)} &= \gamma_{K\,f_b}^{(100)} = 2\,C_{F_N}\,\qquad
  \gamma_{K\,A_N}^{(100)} = 2\,C_{A_N}\,\qquad
  \gamma_{K\,f_m}^{(100)} = \gamma_{K\,f_l}^{(100)} = 
  \gamma_{K\,A_M}^{(100)} = \gamma_{K\,A_U}^{(100)} = 0\\
  \gamma_{K\,f_m}^{(010)} &= \gamma_{K\,f_b}^{(010)} = 2\,C_{F_M}\,\qquad
  \gamma_{K\,A_M}^{(010)} = 2\,C_{A_M}\,\qquad
  \gamma_{K\,f_n}^{(010)} = \gamma_{K\,f_l}^{(010)} = 
  \gamma_{K\,A_N}^{(010)} = \gamma_{K\,A_U}^{(010)} = 0\\
  \gamma_{K\,f_y^i}^{(001)} &= 2\,Q^2_{f_y^i}\ (y\in\{l,m,n,b\})\qquad
  \gamma_{K\,A_N}^{(001)} = \gamma_{K\,A_M}^{(001)} = \gamma_{K\,A_U}^{(001)} = 0\\
  \gamma_{K\,x}^{(200)} &= \frac{K^{(200)}}{2}\gamma_{K\,x}^{(100)}\,,\qquad
  K^{(200)} = C_{A_N}\Lx\frac{2}{3}-\zeta_2\Rx + \frac{10}{3}\betaN{200}\\
  \gamma_{K\,x}^{(020)} &= \frac{K^{(020)}}{2}\gamma_{K\,x}^{(010)}\,,\qquad
  K^{(020)} = C_{A_M}\Lx\frac{2}{3}-\zeta_2\Rx + \frac{10}{3}\betaM{020}\\
  \gamma_{K\,x}^{(002)} &= \frac{K^{(002)}}{2}\gamma_{K\,x}^{(001)}\,,\qquad
  K^{(002)} = \frac{10}{3}\betaU{002}\\
  \gamma_{K\,x}^{(110)} &= \gamma_{K\,x}^{(101)} = \gamma_{K\,x}^{(011)} = 0\ 
      (x\in\{f_l,f_n,f_m,f_b,A_N,A_M,A_U\}\,.
\end{split}
\end{equation}

\subsection{The $\cg$ Anomalous Dimensions}
\begin{equation}
\begin{split}
  \cg_{f_n}^{(100)} &= \cg_{f_b}^{(100)} = \frac{3}{2}\,C_{F_N}
      + \frac{\ep}{2}C_{F_N}\Lx8-\zeta_2\Rx\,\qquad
  \cg_{A_N}^{(100)} = 2\,\betaN{200} - \frac{\ep}{2}C_{A_N}\,\zeta_2\,\qquad
  \cg_{f_{m,l}}^{(100)}=  \cg_{A_{M,U}}^{(100)} = 0\\
  \cg_{f_m}^{(010)} &= \cg_{f_b}^{(010)} = \frac{3}{2}\,C_{F_M}
      + \frac{\ep}{2}C_{F_M}\Lx8-\zeta_2\Rx\,\qquad
  \cg_{A_M}^{(010)} = 2\,\betaM{020} - \frac{\ep}{2}C_{A_M}\,\zeta_2\,\qquad
  \cg_{f_{n,l}}^{(010)} = \cg_{A_{N,U}}^{(010)} = 0\\
  \cg_{f_x^i}^{(001)} &= \frac{3}{2}\,Q^2_{f_x^i}
      + \frac{\ep}{2}Q^2_{f_x^i}\Lx8-\zeta_2\Rx\ (x\in\{l,m,n,b\})\,\qquad
  \cg_{A_U}^{(001)} = 2\,\betaU{002}\,\qquad
  \cg_{A_{M,N}}^{(001)} = 0\\
  \cg_{f_n}^{(200)} &= \cg_{f_b}^{(200)} = 
     C_{F_N}^2\Lx\frac{3}{16} - \frac{3}{2}\zeta_2 + 3\zeta_3\Rx
   + C_{F_N}\,\betaN{200}\Lx\frac{209}{36} + \zeta_2\Rx
   + C_{F_N}\,C_{A_N}\Lx\frac{41}{72} - \frac{13}{4}\zeta_3\Rx\\
  \cg_{A_N}^{(200)} &= 2\,\betaN{300}
   + C_{A_N}\,\betaN{200}\Lx\frac{19}{18}-\zeta_2\Rx
   + C_{A_N}^2\Lx\frac{177}{216} - \frac{1}{4}\zeta_3\Rx\,\qquad
  \cg_{f_{m,l}}^{(200)} = \cg_{A_{M,U}}^{(200)} = 0\\
  \cg_{f_m}^{(020)} &= \cg_{f_b}^{(020)} = 
     C_{F_M}^2\Lx\frac{3}{16} - \frac{3}{2}\zeta_2 + 3\zeta_3\Rx
   + C_{F_M}\,\betaM{020}\Lx\frac{209}{36} + \zeta_2\Rx
   + C_{F_M}\,C_{A_M}\Lx\frac{41}{72} - \frac{13}{4}\zeta_3\Rx\\
  \cg_{A_M}^{(020)} &= 2\,\betaM{030}
   + C_{A_M}\,\betaM{020}\Lx\frac{19}{18}-\zeta_2\Rx
   + C_{A_M}^2\Lx\frac{177}{216} - \frac{1}{4}\zeta_3\Rx\,\qquad
  \cg_{f_{n,l}}^{(020)} = \cg_{A_{N,U}}^{(020)} = 0\\
  \cg_{f_x^i}^{(002)} &= 
     Q^4_{f_x^i}\Lx\frac{3}{16} - \frac{3}{2}\zeta_2 + 3\zeta_3\Rx
   + Q^2_{f_x^i}\,\betaU{002}\Lx\frac{209}{36} + \zeta_2\Rx\,\qquad
  \cg_{A_U}^{(002)} = 2\,\betaU{003}\,\qquad
  \cg_{A_{N,M}}^{(002)} = 0\\
  \cg_{f_b}^{(110)} &= 
     C_{F_N}\,C_{F_M}\Lx\frac{3}{16} - \frac{3}{2}\zeta_2 + 3\zeta_3\Rx\,\qquad
  \cg_{A_N}^{(110)} = 2\,\betaN{210}\,\qquad
  \cg_{A_M}^{(110)} = 2\,\betaM{120}\,\qquad
  \cg_{f_{n,m,l}}^{(110)} = \cg_{A_U}^{(110)} = 0\\
  \cg_{f_{\{b,n\}}^i}^{(101)} &= 
     C_{F_N}\,Q^2_{f_{\{b,n\}}^i}\Lx\frac{3}{16} - \frac{3}{2}\zeta_2 + 3\zeta_3\Rx\ \quad
  \cg_{A_N}^{(101)} = 2\,\betaN{201}\,\qquad
  \cg_{A_U}^{(101)} = 2\,\betaU{102}\,\qquad
  \cg_{f_{m,l}}^{(101)} = \cg_{A_M}^{(101)} = 0\\
  \cg_{f_{\{b,m\}}^i}^{(011)} &= 
     C_{F_M}\,Q^2_{f_{\{b,m\}}^i}\Lx\frac{3}{16} - \frac{3}{2}\zeta_2 + 3\zeta_3\Rx\,\quad
  \cg_{A_M}^{(011)} = 2\,\betaM{021}\,\qquad
  \cg_{A_U}^{(011)} = 2\,\betaU{012}\,\qquad
  \cg_{f_{n,l}}^{(011)} = \cg_{A_N}^{(011)} = 0\,.
\end{split}
\end{equation}

\end{document}